\documentclass[12pt]{article}
\usepackage{epsfig}
\usepackage{comment}
\usepackage{latexsym}
\usepackage{color}
\newcommand{\mysquare}[0]{\raise-.2ex\hbox{{\Large$\Box$}}}
\def\lsim{\mathrel{\rlap {\raise.5ex\hbox{$ < $}}
{\lower.5ex\hbox{$\sim$}}}}
\def\gsim{\mathrel{\rlap {\raise.5ex\hbox{$ > $}}
{\lower.5ex\hbox{$\sim$}}}} \topmargin -1.5cm \textheight=22.5cm
\catcode`\@=11
\newcount\hour
\newcount\minute
\newtoks\amorpm
\hour=\time\divide\hour by60 \minute=\time{\multiply\hour by60
\global\advance\minute by-\hour}
\edef\standardtime{{\ifnum\hour<12 \global\amorpm={am}%
        \else\global\amorpm={pm}\advance\hour by-12 \fi
        \ifnum\hour=0 \hour=12 \fi
        \number\hour:\ifnum\minute<10 0\fi\number\minute\the\amorpm}}
\edef\militarytime{\number\hour:\ifnum\minute<10 0\fi\number\minute}
\def\draftlabel#1{{\@bsphack\if@filesw {\let\thepage\relax
   \xdef\@gtempa{\write\@auxout{\string
      \newlabel{#1}{{\@currentlabel}{\thepage}}}}}\@gtempa
   \if@nobreak \ifvmode\nobreak\fi\fi\fi\@esphack}
        \gdef\@eqnlabel{#1}}
\def\@eqnlabel{}
\def\@vacuum{}
\def\draftmarginnote#1{\marginpar{\raggedright\scriptsize\tt#1}}
\def\draft{\oddsidemargin -.2truein
        \def\@oddfoot{\sl preliminary draft \hfil
        \rm\thepage\hfil\sl\today\quad\militarytime}
        \let\@evenfoot\@oddfoot \overfullrule 3pt
        \let\label=\draftlabel
        \let\marginnote=\draftmarginnote
   \def\@eqnnum{(\theequation)\rlap{\k

 ern\marginparsep\tt\@eqnlabel}%
\global\let\@eqnlabel\@vacuum}  }

\newcommand{\be}[0]{\begin{equation}}
\newcommand{\ee}[0]{\end{equation}}
\newcommand{\ba}[0]{\begin{eqnarray}}
\newcommand{\ea}[0]{\end{eqnarray}}
\def\bs{\begin{subequations}}
\def\es{\end{subequations}}
\def\d{\partial}

\def\thebibliography#1{%
\vskip 0.5cm \centerline{\bf \Large References}
\list{%
[\arabic{enumi}]}{\settowidth\labelwidth{[#1]}
\leftmargin\labelwidth \advance\leftmargin\labelsep
\usecounter{enumi}}
\def\newblock{\hskip .11em plus .33em minus .07em}
\sloppy\clubpenalty4000\widowpenalty4000 \sfcode`\.=1000\relax}

\renewcommand{\theequation}{\arabic{section}.\arabic{equation}}

\renewcommand{\section}{\setcounter{equation}{0}\@startsection
{section}{1}{0mm}{-\baselineskip}{0.5\baselineskip}
{\normalfont\Large\bfseries}}

\renewcommand{\subsection}{\@startsection
{subsection}{2}{0mm}{-\baselineskip}{0.5\baselineskip}
{\normalfont\large\bfseries}}

\renewcommand{\subsubsection}{\@startsection
{subsubsection}{3}{0mm}{-\baselineskip}{0.5\baselineskip}
{\normalfont\normalsize\slshape}}

\usepackage{amssymb,amsfonts}
\usepackage{graphicx}
\usepackage{cite}

\def\bc{\begin{center}}
\def\ec{\end{center}}
\def\bea{\begin{eqnarray}}
\def\eea{\end{eqnarray}}

\def\d{\delta}
\def\t{\tau}
\def\l{\lambda}
\def\L{\Lambda}

\def\k{\kappa}

\def\ve{\varepsilon}

\def\D{\Delta}
\def\taut{\tilde \tau}
\def\ttau{\tilde \tau}

\def\A{{\cal A}}

\def\tt{\tilde t}
\def\hk{\hat k}
\def\dm{\d_M^2}
\def\dt{\d^2_T}                      
\def\tve{{\tilde \varepsilon}}
\def\hi{{\hat \imath}}
\def\hn{{\hat n}}
\def\ti{{\tilde \imath}}

\newcommand{\ie}{{\em i.e. }}

\newcommand{\intN}{\mathbb{N}}
\newcommand{\R}{\mathbb{R}}

\newcommand{\Z}{\mathbb{Z}}

\newcommand{\N}{{\cal N}}

\begin{document}

\begin{titlepage}
\begin{flushright}
LPTENS--07/22, CPHT--RR025.0407\\
June 2007
\end{flushright}

\vspace{0mm}

\begin{centering}
{\bf\huge Inflationary de Sitter Solutions}\\
\vspace{2mm}
{\bf\huge from Superstrings$^\ast$}\\

\vspace{4mm}
 {\Large Costas~Kounnas$^{1}$
and Herv\'e~Partouche$^2$}

\vspace{2mm}

$^1$ Laboratoire de Physique Th\'eorique,
Ecole Normale Sup\'erieure,$^\dagger$ \\
24 rue Lhomond, F--75231 Paris cedex 05, France\\
{\em Costas.Kounnas@lpt.ens.fr}

\vspace{2mm}

$^2$ {Centre de Physique Th\'eorique, Ecole
Polytechnique,$^\diamond$
\\
F--91128 Palaiseau, France\\
{\em Herve.Partouche@cpht.polytechnique.fr}}

\vspace{4mm}

{\bf\Large Abstract}

\end{centering}



\indent In the framework of superstring compactifications with $N=1$
supersymmetry spontaneously broken, (by either geometrical
fluxes, branes or else), we show the existence of \emph{new
inflationary solutions}. The time-trajectory of the scale factor  of
the metric $a$, the supersymmetry  breaking scale $m\equiv m(\Phi)$ and the
temperature $T$ are such that $a \, m $ and $a\, T$ remain
constant.
These solutions request the presence of special
moduli-fields:\\
$i)$ The universal ``no-scale-modulus'' $\Phi$, which
appears in all $N=1$ effective supergravity theories and defines the
supersymmetry breaking scale $m(\Phi)$.\\
$ii)$ The modulus $\Phi_s$, which appears in \emph{a very large class of string
compactifications} and has a $\Phi$-dependent kinetic term. During
the time evolution, $a^4\rho_s$ remains constant as well, ($\rho_s$
being the energy density induced by the motion of $\Phi_s$).

The cosmological term $\Lambda(am)$, the curvature term $k(am, a
T)$ and the radiation term $c_R=a^4\,\rho$ are dynamically generated in a
\emph{controllable way} by radiative and temperature corrections; they
are effectively constant during the time evolution.

Depending on $\Lambda$, $k$ and $c_R$, either a first or second order phase transition can occur in the cosmological scenario. In the first case, an instantonic Euclidean solution exists and connects via tunneling the
inflationary evolution  to another cosmological branch.
The latter starts with a big bang and, in the case the transition does not
occur, ends with a big crunch. In the second case, the big bang and the inflationary phase are smoothly connected.

\vspace{5pt} \vfill \hrule width 6.7cm \vskip.1mm{\small \small\small \noindent 
$^\ast$\ Research partially supported by the EU (under the contracts MRTN-CT-2004-005104, MRTN-CT-2004-512194, MRTN-CT-2004-503369, MEXT-CT-2003-509661), INTAS grant 03-51-6346, CNRS PICS 2530, 3059 and 3747,  and ANR (CNRS-USAR) contract  05-BLAN-0079-01.\\
$^\dagger$\ Unit{\'e} mixte  du CNRS et de l'Ecole Normale
Sup{\'e}rieure associ\'ee \`a l'Universit\'e Pierre et Marie Curie (Paris
6), UMR 8549.\\
 $^\diamond$\ Unit{\'e} mixte du CNRS et de l'Ecole Polytechnique,
UMR 7644.}

\end{titlepage}
\newpage
\setcounter{footnote}{0}
\renewcommand{\thefootnote}{\arabic{footnote}}
 \setlength{\baselineskip}{.7cm} \setlength{\parskip}{.2cm}


\setcounter{section}{0}
\section{Introduction}
\label{intro}

In the framework of superstring and $M$-theory compactifications,
there are always moduli fields coupled
in a very special way to the gravitational and matter sector of the
effective $N=1$ four-dimensional supergravity.
The gravitational and the scalar field
part of the effective Lagrangian have the generic form
 \be
 {\cal L}=
\sqrt{-\det g}\left[{1\over 2}R -g^{\mu\nu}~K_{i \bar \jmath}
 ~\partial_{\mu}\phi_i \partial_{\nu} \bar\phi_{\bar \jmath}~
-V (\phi_i, \bar\phi_{\bar \imath})\right]\, ,
 \ee
where  $K_{i \bar \jmath}$ is the metric of the scalar manifold
and $V$ is the scalar potential of the $N=1$ supergravity.
(We will always work
in gravitational mass units, with
$ M ={1\over \sqrt{ 8 \pi G_N}}=2.4\times10^{18}$ GeV).
What will be crucial in this work is the non-triviality of the scalar
kinetic terms $K_{i \bar \jmath}$ in the $N=1$ effective
supergravity theories that will provide us, in some special cases,
accelerating cosmological solutions once the radiative and
temperature corrections are taken into account.\\
$~$\\
Superstring vacua with spontaneously broken supersymmetry  
 \cite{StringSusyBreaking}  that are
consistent at the classical level with a flat space-time define a
very large class of ``no-scale" supergravity models \cite{noScale}.
Those with $N=1$  spontaneous  breaking deserve more attention. Some
of them are candidates for describing (at low energy) the physics of
the  standard model and extend it up to ${\cal O}(1)$~TeV energy
scale. This class of models contains an enormous number of
consistent string vacua that can be constructed either via freely
acting orbifolds \cite{StringSusyBreaking, freelyOrbifolds} or 
``geometrical fluxes'' \cite{GeometricalFluxes} in heterotic
string and type IIA,B orientifolds, or with non-geometrical
fluxes \cite{Fluxes} (e.g. RR-fluxes or else).\\
$~$\\
 Despite the plethora of this type of vacua,
  an interesting class of them are those which are described by an effective 
  $N=1$ ``no-scale supergravity theory". Namely the vacua in which the supersymmetry is sponaneously broken with a vanishing classical potential
with undetermined gravitino mass due to at least one flat field direction, the ``no-scale modulus $\Phi$.  At the quantum level a non-trivial effective potential is radiatively generated which may or may not stabilize the ``no-scale" 
modulus \cite{noScale}.\\
$~$\\
 What we will explore in this work are the  universal scaling  properties of the ``thermal" effective potential at finite temperature that emerges at the quantum level of the theory. As we will show in section \ref{effpot},  the quantum and thermal corrections are under control, (thanks to supersymmetry and to  
the classical structure of the ``no-scale models"), showing interesting scaling properties. \\
$~$\\
In section \ref{conventions}, we set up our notations and conventions in the effective 
$N=1$ ``no-scale" supergravities of the type IIB orientifolds with $D_3$-branes and non-trivial NS-NS and RR three form fluxes $H^3$ and $F^3$. 
We identify the ``no-scale" modulus $\Phi$, namely the scalar
superpartner of the Goldstino which has the property to couple to the
trace of the energy momentum tensor of a sub-sector of the theory
 \cite{AKcritical}.
 More importantly, it defines the
field-dependence of the gravitino mass \cite{noScale}
 \be
 \label{mphi}
m(\Phi)=e^{\alpha \Phi}.
\ee
Other extra moduli that we will consider here are those
with $\Phi$-dependent kinetic terms. These moduli appear naturally in all string compactifications  \cite{PhiKineticTermes}. 
We are in particular interested in scalars ($\Phi_s$) which are 
leaving on $D_3$-branes  and whose 
 kinetic terms scale as the inverse volume of  the ``no-scale" moduli space.\\
$~$\\
In section \ref{eqmotion}, we display the relevant gravitational, fields and thermal equations of motion in the context of a Friedman-Robertson-Walker (FRW) space-time. We actually generalize the mini-superspace (MSS) action by including fields with non-trivial kinetic terms and a generic, scale factor dependent, thermal effective potential.\\
$~$\\
 In our analysis we restrict ourselves to the large moduli limit, neglecting non-perturbative terms and world-sheet instanton corrections  ${\cal O}(e^{-S})$, ${\cal O}(e^{-T_a})$.    On the other hand we keep the perturbative quantum and thermal corrections.\\
$~$\\
 Although this study looks hopeless and out of any systematic control even at the perturbative level, it turns out to be manageable thanks to the initial no-scale structure appearing at the classical level  (see section \ref{effpot}). \\
$~$\\
In section \ref{crisol}, we show the existence of a critical solution to the equations of motion that follows from the scaling properties derived in section \ref{effpot}. We have to stress here that we extremize the effective action by solving the gravitational and moduli equations of motion and do not consider the stationary solutions emerging from a minimization of the effective potential only. We find in particular that a universal solution exists where all scales evolve in time in a similar way, so that their ratios remain constant: ${m(t)/ T(t)}=$ const., $a(t)\, m(t)=$ const.. Along this trajectory, effective time-independent cosmological term $\L$, curvature term $k$ and radiation term are generated in the MSS action, characterizing the cosmological evolution. \\
$~$\\
Obviously, the validity of the cosmological solutions based on (supergravity) effective field theories is limited. For instance, in the framework of more fundamental theories such as string theory, there are high temperature instabilities occuring at $T\simeq T_H$, where $T_H$ is the Hagedorn temperature of order the mass of the first string excited state. To bypass these limitations, one needs to go beyond the effective field theory approach and consider the full string theory (or brane, M-theory,...) description. Thus, the effective solutions presented in this work are not valid anymore and {\em must be corrected} for temperatures above $T_H$. Moreover, Hagedorn-like instabilities can also appear in general in other corners of the moduli space of the fundamental theory, when space-time supersymmetry is spontaneously broken. \\
$~$\\
Regarding the temperature scale as the inverse radius of the compact Euclidean time, one could conclude that all the internal radii of a higher dimensional fundamental theory have to be above the Hagedorn radius. This would mean that the early time cosmology should be dictated by a 10-dimensional picture rather than a 4-dimensional one where the internal radii are of order the string scale. There is however a loophole in this statement. Indeed, no tachyonic instability is showing up in the whole space of the moduli which are not involved in the spontaneous breaking of supersymmetry, as recently shown in explicit examples \cite{CKPT}. This leeds us to the conjecture that the only Hagedorn-like restrictions on the moduli space depend on the supersymmetry breaking. In our cosmological solutions, not only the temperature $T$ scale is varying, but also the supersymmetry breaking scale $m$, which turns to be a moduli-dependent quantity. Based on the above statements, we expect that in a more accurate stringy description of our analysis, there should be restrictions on the temperature as well as the supersymetry breaking scale. This has been recently explicitly shown in the stringy examples considered in  \cite{CKPT}.\\
$~$\\
In section \ref{othmod}, our cosmological solutions are generalized by  including moduli with other scaling properties of their kinetic terms.\\
$~$\\
Finally, section \ref{perscl} is devoted to our conclusions and perspective for future work. 


\section{N=1 No-Scale Sugra from Type IIB Orientifolds}
\label{conventions}

In the presence of branes and fluxes, several moduli can be stabilized. For instance, in ``generalized''  Calabi-Yau compactifications,  either the $h_{1,1}$
 K\"alher structure  moduli or the $h_{2,1}$ complex structure moduli can be stabilized according to the brane and flux configuration in type IIA or type IIB orientifolds \cite{GeometricalFluxes, Fluxes1, Fluxes, AKcritical}. The (partial) stabilization of the moduli can 
 lead us at the classical level to AdS like solutions, domain wall solutions or 
 ``flat no-scale like solutions". Here we will concentrate our attention on the ``flat no-scale like solutions".
\\
$~$\\
In order to be more explicit, let us
consider as an example the type IIB orientifolds with  $D_3$-branes and
non-trivial NS-NS and RR three form fluxes $H^3$ and $F^3$.
This particular configuration  induces a well known
superpotential $W(S,U_a)$ that can stabilize all complex structure moduli
$U_a$ and the coupling constant modulus $S$ \cite{Fluxes, GeometricalFluxes}. 
The remaining $h_{1,1}$
moduli $T_a$  ``still remain flat directions  at the classical level", 
e.g.  neglecting world-sheet instanton corrections ${\cal O}(e^{-T_a})$ 
and the perturbative and non-perturbative quantum corrections \cite{Fluxes}.\\ 
$~$\\
It is also well known by now that in the large $T_a$ limit  the K\"alher potential is given by the
intersection numbers $d_{abc}$ of the special geometry of the
Calabi-Yau manifold and orbifold compactifications \cite{SpecialGeometry, Veffective}: 
\be K=-\log\, d_{abc}(T_a+\bar T_a)(T_b+\bar T_b)(T_c+\bar T_c)\, . 
\ee 
Thus, after the $S$ and $U_a$ moduli stabilization, 
the superpotential $W$ is effectively constant and implies a vanishing potential in all $T_a$ directions. The gravitino mass term is however non-trivial
 \cite{noScale, StringSusyBreaking, GeometricalFluxes, Fluxes, Veffective},
 \be
 m^2=|W|^2e^{K}\, .
\ee
This classical property of ``no-scale models" emerges from the
cubic form of $K$ in the moduli $T_a$ and is generic in all
type IIB orientifold compactifications with $D_3$-branes and three form $H^3$ and $F^3$ 
fluxes \cite{Fluxes, GeometricalFluxes}.
 Keeping for simplicity the direction $T_a=\gamma_a T$ (for some constants $\gamma_a$) and
freezing all other directions, the K\"alher potential is taking the well known $SU(1,1)$ structure \cite{noScale},
\be
K=-3\log(T+\bar T)\, .
\ee
This gives rise to the kinetic term and gravitino mass term,
\be
g^{\mu\nu}~3{\partial_{\mu} T
\partial_{\nu} \bar T \over (T+\bar T)^2} \quad \mbox{and}\quad
m^2= c e^K ={c\over (T+\bar T)^3}\, ,
\ee
where $c$ is a constant.
Freezing Im$T$ and defining
the field $\Phi$ by
\be
e^{2\alpha \Phi}= m^2={c\over (T+\bar T)^3}\, ,
\ee
one finds a kinetic term
\be
g^{\mu\nu}~3{\partial_{\mu} T\partial_{\nu}
\bar T \over (T+\bar T)^2}= g_{\mu\nu}~{\alpha^2 \over
3}~\partial_{\mu} \Phi\partial_{\nu} \Phi\, .
\ee
The choice $\alpha^2=3/2$ normalizes canonically the kinetic
term of the modulus $\Phi$.\\
$~$\\
The  other extra moduli that we will consider are those
with $\Phi$-dependent kinetic terms.
We are in particular interested to the scalars whose kinetic terms scale as the inverse volume of the $T$-moduli. For one of them, $\Phi_s$, one has 
 \be
K_s \equiv -{\alpha^2 \over
3}~e^{2\alpha \Phi}~ g^{\mu\nu}~\partial_{\mu} \Phi_s\partial_{\nu}
\Phi_s~=-{\alpha^2 c\over
3}~g^{\mu\nu} ~{\partial_{\mu} \Phi_s\partial_{\nu} \Phi_s
\over (T+\bar T)^3}.
\ee
Moduli with this scaling property appear in {\em  a very
large class of string compactifications}. Some examples are: \\
$i)$ All moduli fields leaving in the parallel space of
$D_3$-branes \cite{Fluxes, GeometricalFluxes}.\\
$ii)$ All moduli coming from the twisted sectors of $\Z_3$-orbifold
compactifications in heterotic string \cite{PhiKineticTermes}, {\em 
after non-perturbative stabilization of $S$ by  gaugino condensation and
flux-corrections} \cite{condflux}. \\
$~$\\
Our analysis will also consider other moduli fields with different scaling
properties, namely those with kinetic terms of the form:
\be 
K_w \equiv -{1 \over 2}\,e^{(6-w)\alpha
\Phi}~g^{\mu\nu}\,\partial_{\mu} \phi_w\partial_{\nu} \phi_w\, , 
\ee
with weight $w=0,2$ and $6.$ 


\section{Gravitational, Moduli and Thermal Equations}
\label{eqmotion}

In a fundamental theory, the number of degrees of
freedom is important (and actually infinite in the
context of string or M-theory). However, in an effective field theory,
an ultraviolet cut-off set by the underlying theory determines
the number of states to be considered. We focus on cases where
these states  include the scalar moduli fields $\Phi$ and $\Phi_s$,
with non-trivial kinetic terms given by
 \be
 \label{lag}
{\cal L}= \sqrt{-\det g}
\left[{1\over 2}R -{1\over 2}~g^{\mu\nu}\left(
\partial_{\mu}\Phi \partial_{\nu}\Phi +
e^{2\alpha\Phi}~\partial_{\mu}\Phi_s
\partial_{\nu}\Phi_s\right)-V(\Phi,\mu)\right] +\cdots
\ee
In this Lagrangian, the ``$\cdots$"  denote all the other degrees of
freedom, while the effective potential $V$ depends on $\Phi$ and the
renormalization scale $\mu$. We are looking for gravitational solutions
based on isotropic and homogeneous FRW space-time metrics,
\be
\label{metric}
ds^2=- N(t)^2\,dt^2 + a(t)^2d\Omega_3^2 \, ,
\ee
where $\Omega_3$ is a
3-dimensional compact  space with constant curvature $k$, such as a
sphere or an orbifold of hyperbolic space. This defines an effective one dimensional
action, the so called
``mini-super-space" (MSS) action \cite{WFU, MSSThermal, radiation-density, KPThermal}.\\
$~$\\
A way to include into the MSS action the
quantum fluctuations of the full metric and matter degrees of
freedom (and thus taking into account the back-reaction on the
space-time metric), is to switch on a thermal bath at
temperature $T$ \cite{MSSThermal, radiation-density, KPThermal}. In this way, the
remaining degrees of freedom are parameterized by a pressure
$P(T,m_i)$ and a density $\rho(T,m_i)$, where $m_i$ are the non-vanishing masses of the theory. Note that $P$ and $\rho$ have an implicit dependence on $\Phi$, through the mass $m(\Phi)$
defined in eq. (\ref{mphi}) \cite{AKcritical}. The
presence of the thermal bath modifies the effective MSS action,
including the corrections due to the quantum fluctuations of the
degrees of freedom whose masses are below the temperature scale $T$. The result,
together with the fields $\Phi$ and $\Phi_s$, reads 
\be \label{act}
S_{\mbox{\tiny \em{eff}}} =-{|k|^{-{3\over 2}}\over 6} \int dt~a^3\left({3\over N}\left( {\dot a \over
a}\right)^2-{3kN \over a^2}-{1\over2N} \dot\Phi^2 -{1\over
2N}~e^{2\alpha \Phi} \dot \Phi_s^2  +N V-{1\over 2N}(\rho+P)+{N\over
2}(\rho-P)  \right),
\ee
where a ``dot" denotes a time derivation.
$N(t)$ is a gauge dependent function that can be
arbitrarily chosen by a redefinition of time. We will always
use the gauge  $N\equiv 1$, unless it is  explicitly specified.\\
$~$\\
The variation with respect to $N$ gives rise to the Friedman equation,
 \be
 \label{Friedmaneq}
3H^2=  -{3k\over a^2}+ \rho+{1\over2} \dot\Phi^2 +{1\over
2}~e^{2\alpha \Phi} \dot \Phi_s^2 +V\, ,
\label{Friedman}
\ee
where $H=\left({\dot a /a}\right)$.\\
$~$\\
The other gravitational equation is obtained
by varying the action with respect to the
scale factor $a$:
\be
 2\dot H +3H^2 =-{k\over a^2}-P-{1\over2}\, 
\dot\Phi^2-{1\over 2}\, e^{2\alpha \Phi} \dot \Phi_s^2 +V +{1\over 3}\, a{\partial
V \over\partial a}\, .
\label{vara}
\ee
In the literature, the last term
$a(\partial V /\partial a)$ is frequently taken to be zero.
However, this is not  valid due to the dependence of $V$ on $\mu$,
when this scale is chosen appropriately as will be seen in section \ref{eqmotion}.
We thus keep this term and will see that it plays a crucial role in the derivation
of the inflationary solutions under investigation.\\
$~$\\
We find useful to replace eq. (\ref{vara}) by the linear sum of
eqs. (\ref{Friedman}) and (\ref{vara}), so that the kinetic terms of $\Phi$ and $\Phi_s$ drop out,
\be
\label{gravityeq}
\dot H +3H^2 =-{2k\over a^2}+{1\over 2}(\rho-P) +V +
{1\over 6}\, a{\partial V \over\partial a}\, .
\label{gravPotential}
\ee
The other field equations are the moduli ones,
\be
\label{Phi}
\ddot\Phi + 3H \dot \Phi +{\partial \over  \partial
\Phi}\left(V-P-{1\over 2}~e^{2\alpha \Phi} \dot \Phi_s^2\right)=0
\ee
and
\be
\label{Phis}
 \ddot \Phi_s + (3H +2 \alpha \dot\Phi )~\dot\Phi_s =0\, .
\ee
The last equation (\ref{Phis}) can be solved
immediately,
\be
\label{Ks}
K_s~\equiv~ {1\over 2}~e^{2\alpha \Phi} \dot
\Phi_s^2~ =~ C_{s}{e^{-2\alpha \Phi}\over a^6}\, ,
\ee
where $C_s$ is a positive  integration constant.
It is important to stress here that we insist to
keep in eq. (\ref{Phi})
 both terms $\partial P/  \partial
\Phi $ and $\partial K_s / \partial \Phi$ that are however
usually omitted in the literature. The first term vanishes
{\em only} under the assumption that all masses are taken to be
$\Phi$-independent, while the absence of the second term assumes
a trivial kinetic term. However, {\em  both assumptions are not valid
in string effective supergravity theories !} (See section \ref{eqmotion}.)\\
$~$\\
{F}inally, we display for completeness the total energy conservation
of the system,
 \be
{d\over dt}\left(\rho +{1\over 2}\dot \Phi ^2 +K_s+
V\right)+3H\left(\rho+P +\dot \Phi ^2 +2K_s\right)=0\, . \ee Before
closing this section, it is useful to derive some extra useful
formulas that are associated to the thermal system. The integrability
condition of the second law of thermodynamics reaches, for the
thermal quantities $\rho$ and $P$,
 \be
T\,{\partial
P \over\partial T}=~\rho +P\, .
\label{entropy}
\ee
The fact that these quantities are four-dimensional implies
\be
\left( m_i{\partial \over \partial m_i} +T {\partial \over
\partial T}\right)\rho =  4\rho \quad \mbox{and}
\quad \left( m_i{\partial \over \partial m_i} +T {\partial \over
\partial T}\right)P =  4P\, .
 \label {scale}
\ee
Then,  the second eq. (\ref{scale}) together with the eq. (\ref{entropy})
implies \cite{AKcritical}:
\be
m_i{\partial P\over
\partial m_i}=-(\rho-3P)\, .
\label{derivp}
\ee
Among the non-vanishing $m_i$, let us denote with ``hat-indices'' the masses $m_{\hi}$
that are $\Phi$-independent, and with ``tild-indices''
the masses $m_{\ti}$ that have the following  $\Phi$-dependence:
\be
\{ m_i \} =\{m_\hi\} \cup \{m_\ti \} \quad \mbox{where}
\quad m_\ti={ c}_{\ti}~e^{\alpha \Phi}\, ,
\label{Phimasses}
\ee
for some constants $c_{\ti}$.
Then, utilizing eq. (\ref{derivp}), we obtain a
very fundamental equation involving the modulus
field $\Phi$ \cite{AKcritical},
\be
\label{r-3p}
-{\partial P\over
\partial \Phi} = \alpha\, (\tilde\rho-3\tilde P)\, ,
\label{pDerivPhi}
\ee
where $\tilde \rho$ and $\tilde P$ are the contributions to
$\rho$ and $P$ associated to the states with $\Phi$-dependent masses $m_\ti$.
The above equation (\ref{pDerivPhi}) clearly shows that the
modulus field $\Phi$ couples to the (sub-)trace of the energy
momentum tensor associated to the thermal system \cite{AKcritical}
$\tilde \rho$, $\tilde P$  of the
states with $\Phi$-dependent masses defined in eq. (\ref{Phimasses}). We
return to this point in the next section.


\section{Effective Potential and Thermal Corrections }
\label{effpot}

In order to find solutions to the coupled gravitational and  moduli
equations discussed in the previous section, it is necessary to
analyze the structure of the scalar potential $V$ and the thermal
functions $\rho$, $P$. More precisely, we have to specify their
dependence on $\Phi, T, a$ and $\Phi_s$.  Although this analysis
looks hopeless in a generic field theory, it is perfectly under
control in the string
effective no-scale supergravity theories.\\
$~$\\
Classically the potential $V_{\mbox{\tiny cl}}$ is zero along the moduli directions
$\Phi$ and $\Phi_s $.
At the quantum level, it receives radiative and thermal corrections
that are given in
terms of the effective potential \cite{Veffective},
$V(m_i,\mu)$, and in terms of the thermal function, $-P(T, m_i)$. Let us consider both types of corrections. 


\subsection{Effective Potential}
\label{effpotential}

The one loop effective potential has the usual form 
\cite{Veffective, StrM2}, \be V= V_{\mbox{\tiny cl}}+{1\over
64\pi^2}{\rm Str} {\cal M}^0 \Lambda_{\mbox{\tiny co}}^4
\log{\Lambda_{\mbox{\tiny co}}^2 \over
  \mu^2} +{1\over 32\pi^2}{\rm Str} {\cal M}^2\Lambda_{\mbox{\tiny co}}^2 +
{1\over 64\pi^2}{\rm Str} \left({\cal M}^4\log{{\cal M}^2 \over
  \mu^2}\right) + \cdots\, ,
\label{Veffective}
\ee
where $V_{\mbox{\tiny cl}}$ is the classical part, which vanishes
in the string effective ``no-scale''
supergravity case. An  ultraviolet
cut-off $\Lambda_{\mbox{\tiny co}}$ is introduced
and $\mu$ stands for the renormalization scale.
\be
{\rm Str} {\cal M}^n\equiv \sum_I (-)^{2J_I}(2J_I+1)~m_I^n
\ee
is a sum over the
$n$-th power of the mass eigenvalues. In our notations,
the index $I$ is referring to both massless and massive states
(with eventually $\Phi$-dependant masses).The weights account
for the numbers of degrees of freedom and the
statistics of the spin $J_I$ particles.\\
$~$\\
The quantum corrections to the vacuum energy
with the highest degree of ultraviolet divergence
is the $\Lambda_{\mbox{\tiny co}}^4$ term,
whose coefficient ${\rm Str} {\cal M}^0=(n^B-n^F)$ is equal
to the number of bosonic minus fermionic degrees of freedom.
This term is thus always absent in supersymmetric
theories since they possess equal numbers of bosonic
and fermionic states.\\
$~$\\
The second most divergent term in eq. (\ref{Veffective}) is the
$\Lambda_{\mbox{\tiny co}}^2$ contribution proportional to ${\rm
Str} {\cal M}^2$. In the $N=1$ spontaneously broken supersymmetric
theories, it is always proportional to the square of the gravitino
mass-term $m(\Phi)^2$, 
\be 
{\rm Str} {\cal M}^2 =  c_2~m(\Phi)^2 \, .
\ee 
The coefficient $ c_2$ is a field independent number. It depends
only on the geometry of the kinetic terms of the  scalar and gauge
manifold, and {\em  not on the details of the
superpotential} \cite{StrM2, Veffective}. This property is
very crucial in our considerations.\\
$~$\\
The last term has a logarithmic behavior with
 respect to the infrared 
scale $\mu$ and is independent of  the ultraviolet cut-off 
$\Lambda_{\mbox{\tiny co}}$. Following the infrared regularization
method valid in string theory  (and field theory as well) adapted in
ref. \cite{StringInfrared}, the scale $\mu$ is proportional to the
curvature of the three dimensional space, \be \mu ={1\over \gamma
a}\, , \label{mu-a} \ee where $ \gamma$ is a numerical coefficient
chosen appropriately according to the renormalization group equation
arguments. Another physically equivalent choice for $\mu$ is to be
proportional to the temperature scale, $\mu=\zeta T$. The curvature
choice (\ref{mu-a}) looks more natural and has the advantage to be
valid even in the
absence of the thermal bath.\\
$~$\\
Modulo the logarithmic term, the ${\rm Str} {\cal M}^4$ can be
expanded in powers of gravitino mass $m(\Phi)$,
\be
{1\over 64 \pi^2}
{\rm Str} {\cal M}^4 = C_4 m^4+C_2m^2 + C_0 \, .
 \ee
Including the logarithmic terms and adding the quadratic contribution
coming from the ${\rm Str} {\cal M}^2$,  we obtain the following
expression for the effective potential organized in powers of
$m(\Phi)$:
\be
V= V_4(\Phi,a) +V_2(\Phi,a) + V_0(\Phi,a)\, ,
\ee
where
\be
\label{vn}
V_n(\Phi,a) = m^n(\Phi)\left(\, C_n+Q_n \log \left(
m(\Phi)\gamma a\right) \phantom{\dot \phi}\!\!\! \!\right)\, ,
\ee
for constant coefficients $C_n$ and $Q_n$, ($n=4,2,0$).
These contributions satisfy
\be
\label{derV}
{\partial{V}_n(\Phi,a)\over
\partial \Phi}=\alpha(\, nV_n + ~m^nQ_n)
\quad \mbox{and}\quad a{\partial{V}_n(\Phi,a)\over
\partial a}=m^nQ_n\, .
\ee 
The logarithmic dependence in the effective potential can be
derived in the effective field theory by considering the
Renormalization Group Equations (RGE). They involve the gauge couplings,
the Yukawa couplings and the soft-breaking terms 
\cite{RGEsoft, Veffective}. These soft-breaking terms are usually parameterized by  the gaugino mass terms
$M_{1/2}$, the soft scalar masses $m_0$, the trilinear coupling mass
term $A$ and the analytic mass term $B$, 
\cite{RGEsoft, Veffective}. However, what will be of main importance in this
work is that {\em  all soft breaking mass terms are proportional
to $m(\Phi)$} \cite{StrM2, Veffective}.


\subsection{Thermal Potential}
\label{thpotential}

For bosonic (or fermionic) fluctuating states of masses $m_b$ (or
$m_f$) in thermal equilibrium at temperature $T$, the general
expressions of the energy density $\rho$ and pressure $P$ are
\be
\label{rP}
\rho =T^4 \left( \sum _{\mbox{\scriptsize boson }b}
I_\rho^B\left({m_b\over T}\right) + \sum _{\mbox{\scriptsize fermion }f}
I_\rho^F\left({m_f\over T}\right)
\right), \,P = T^4 \left( \sum _{\mbox{\scriptsize boson }b}
I_P^B\left({m_b\over T}\right) + \sum _{\mbox{\scriptsize fermion }f}
I_P^F\left({m_f\over T}\right) \right) \, ,
\end{equation}
where
\be
\label{i}
I_\rho^{B(F)}\left({m\over T}\right)=
\int_0^\infty dq {q^2 E\left(q,{m\over T}\right) \over
  e^{E\left(q,{m\over T}\right)}\mp1} \, , \quad
I_P^{B(F)}= {1\over 3} \int_0^\infty dq {q^4 /E\left(q,{m\over T}\right) \over
e^{E\left(q,{m\over T}\right)} \mp1 }
\ee
 and
$E\left(q,{m\over T}\right) = \sqrt{q^2+{m^2\over T^2}}$. \\
$~$\\
There are three distinct sub-sectors of states:\\
$i)$ The sub-sector of $n_0^B$ bosonic and  $n_0^F$ fermionic massless states. From eqs. (\ref{rP}) and (\ref{i}), their  energy density $\rho_0$ and pressure $P_0$ satisfy 
\begin{equation}
\label{rhoT4}
\rho_0 = 3 P_0={\pi^4\over 15}\left(n_{0}^B+{7\over
8}n_{0}^F\right)  T^4\, .
\end{equation}
In particular, we have
 $\rho_{0}-3P_{0}=0$ and $\partial P_{0}/\partial \Phi= 0$.\\
$ii)$  The sub-sector of states with non vanishing
masses {\em independent} of $m({\Phi})$.\\ 
\indent {\scriptsize $\bullet$} Consider the $\hn_0^B$ bosons and $\hn_0^F$ fermions whose masses we denote by $m_{\hi_0}$ are below T. The energy density $\hat \rho$ and pressure $\hat P$ associated to them satisfy
\be
\hat \rho(T,m_{\hi_0})=\hat \rho(T, m_{\hi_0}=0) +
m^2_{\hi_0}{\partial \hat \rho\over \partial m^2_{\hi_0}} =
{\pi^4 \over 15}\left(\hn_0^B+{7\over
8}\hn_0^F\right)  ~T^4 -\sum_{\hi_0}
{\hat c}_{\hi_0}~m_{\hi_0}^2~T^2\, ,
\ee
\be
\hat P(T,m_{\hi_0})=\hat P(T, m_{\hi_0}=0) +
m^2_{\hi_0}{\partial \hat P\over \partial m^2_{\hi_0}} =
{\pi^4 \over 45}\left(\hn_0^B+{7\over
8}\hn_0^F\right)  ~T^4 -\sum_{\hi_0}
{\hat c}_{\hi_0}~m_{\hi_0}^2~T^2\, ,
\ee
where the ${\hat c}_{\hi_0}$'s are non-vanishing positive constants. In particular, one has $\partial \hat P/\partial \Phi= 0$.\\
\indent {\scriptsize $\bullet$} For the masses  $m_{\hat \imath}$ above $T$, the
contributions of the particular degrees of freedom are
exponentially suppressed and decouple from the thermal system. We are not including their contribution. \\
$iii)$ The sub-sector with non vanishing masses {\em
proportional to} $m({\Phi})$. Its energy density $\tilde \rho$ and pressure $\tilde P$ satisfy
\be
\label{derP}
{\partial P\over \partial \Phi}=-\alpha\, (\tilde \rho-3\tilde P)\, ,
 \ee
 as was shown at the end of section \ref{conventions}. 
This identity is also valid for the massless system we
consider in case $i)$.\\
$~$\\
According to the scaling behaviors with respect to $T$
and $m(\Phi)$, we can separate
\be
\label{rP42}
\rho=\, \rho_4 ~+~\rho_2\; ,~~~~~~~P=\, P_4~+~P_2 \, ,
\ee
where
\be
\left(\, m(\Phi)~{\partial \over \partial
m(\Phi)}~+~T~{\partial \over \partial T~} \right)~(\rho_n,~P_n)=
n~(\rho_n,~ P_n)\, .
\ee
$\rho_4$ and $P_4$ are the sums of the contributions of the massless
states (case $i)$), the $T^4$ parts of $\hat \rho$ and $\hat P$
(case $ii)$), and $\tilde \rho$ and $\tilde P$ (case $iii)$),
\be
\label{r4}
\rho_4=T^4\left( {\pi^4\over 15}\left(\left(n_0^B + \hn_0^B\right)+{7\over 8}\left(n_0^F + \hn_0^F\right)\right) +
\sum _{\mbox{\scriptsize boson }\tilde b}I^B_\rho\left({m_{\tilde b}\over T}\right) +
\sum _{\mbox{\scriptsize fermion }\tilde f}I^F_\rho\left({m_{\tilde f}\over T}\right) \right)\, ,
\ee
\be
\label{P4}
P_4=T^4\left( {\pi^4\over 45}\left(\left(n_0^B + \hn_0^B\right)+{7\over 8}\left(n_0^F + \hn_0^F\right)\right)+
\sum_{\mbox{\scriptsize boson }\tilde b}I^B_P\left({m_{\tilde b}\over T}\right) +
\sum_{\mbox{\scriptsize fermion }\tilde f}I^F_P\left({m_{\tilde f}\over T}\right) \right)\, ,
\ee
while $\rho_2$ and $P_2$ arise from the $T^2$ parts of $\hat \rho$ and
$\hat P$ (case $ii)$):
\be
\label{T2}
\rho_2=P_2=-\sum_{\hi_0} \hat c_{\hi_0}~m_{\hi_0}^2~T^2\equiv
-{\hat M^2}~T^2 \, .
\ee


\section{Critical Solution}
\label{crisol}

The fundamental ingredients in our
analysis are the scaling properties
of the total effective potential at
finite temperature,
\be
V_{\mbox{\tiny total}}=V-P\, .
\ee
Independently of the complication
appearing in the radiative and temperature corrected effective potential,  the scaling violating terms are  under
control. Their structure suggests to search for
a solution where all the scales of
the system, $m(\Phi)$, $T$ and $\mu=(1/\gamma a)$,
remain proportional during their evolution in time,
\be
\label{anzats}
e^{\alpha\Phi}\equiv m(\Phi)={1\over {\gamma'} a }~~\Longrightarrow
~~H=-{\alpha} \dot \Phi\qquad \mbox{and}\qquad {\xi} ~{m(\Phi)}\,=\,T \, .
\ee
Our aim is thus to determine the constants $\gamma'$ and $\xi$  in terms of $C_s$ in
eq. (\ref{Ks}), $\gamma$, and the computable quantities $C_n$, $Q_n$, $(n=4,2,0)$
in string theory, such that the equations of motion for $\Phi$,
$\Phi_s$ and the gravity are satisfied. On the trajectory (\ref{anzats}),
the contributions $V_n$, $(n=4,2,0)$ defined in eq. (\ref{vn}) satisfy
\be
\label{VN}
V_n=m^n C_n'\qquad \mbox{where} \qquad C_n'=
C_n+Q_n\log\left({\gamma\over \gamma'}\right)\, ,
\ee
and
\be
{\partial V_n\over \partial \Phi}=\alpha \, m^n(nC_n'+ Q_n)\; ,
\qquad a\, {\partial V_n\over \partial a}=m^nQ_n\, .
\ee
Also, the contributions of $\Phi$ and $1/a^6$ in $K_s$ in eq. (\ref{Ks})
conspire to give a global $1/a^4$ dependence,
\be
\label{CS1}
K_s= C_s\, {\gamma^{\prime 2}\over a^4}\, .
\ee
 Finally, the sums over the full towers of states
with $\Phi$-dependent masses behave in $\rho_4/T^4$
and $P_4/T^4$ as pure constants, (see eqs. (\ref{r4}) and (\ref{P4})),
\be
\label{r4c}
\rho_4=r_4 T^4\quad \mbox{where}\quad r_4={\pi^4\over 15}\left(\left(n_0^B + \hn_0^B\right)+{7\over 8}\left(n_0^F + \hn_0^F\right)\right)+
\sum_{\tilde b}I^B_\rho\left({\tilde c_{\tilde b}\over \xi}\right) +
\sum_{\tilde f}I^F_\rho\left({\tilde c_{\tilde f}\over \xi}\right)\, ,
\ee
\be
\label{P4c}
P_4=p_4 T^4\quad \mbox{where}\quad p_4={\pi^4\over 45}\left(\left(n_0^B + \hn_0^B\right)+{7\over 8}\left(n_0^F + \hn_0^F\right)\right)+
\sum_{\tilde b}I^B_P\left({\tilde c_{\tilde b}\over \xi}\right) +
\sum_{\tilde f}I^F_P\left({\tilde c_{\tilde f}\over \xi}\right)\, .
\ee
As a consequence,
using eqs. (\ref{derV}) and (\ref{derP}), the $\Phi$-equation (\ref{Phi}) becomes, 
\be
\dot H + 3H^2 =\alpha^2\left(\phantom{\dot\Phi}\!\!\!\!
(4C_4' +Q_4)m^4+(2C_2'+Q_2)m^2+Q_0+
(r_4-3p_4)\xi^4m^4-2C_s\gamma^{\prime 6}m^4
\phantom{\dot\Phi}\!\!\!\!\right).
\ee
On the other hand, using eq. (\ref{derV}), the gravity equation (\ref{gravityeq})
takes the form
\be
\label{gravcri}
\dot H +
3H^2=-2k\gamma^{\prime 2}m^2+ {1\over 2}(r_4-p_4)\xi^4m^4  +
(C'_4m^4+C'_2m^2+C'_0) +{1\over 6}(Q_4m^4 +Q_2m^2 +Q_0)\, .
\ee
The compatibility of the $\Phi$-equation and the gravity equation
along the critical trajectory implies an identification of
the coefficients of the monomials in $m$.
The constant terms determine $C'_0$ in term of $Q_0$
\be
\label{C0'}
C'_0={6\alpha^2-1\over 6}~Q_0\, ,
\ee
which amounts to fixing $\gamma'$,
\be
\gamma'=\gamma \, e^{{C_0\over Q_0}-{6\alpha^2-1\over 6}}\, .
\ee
The quadratic terms determine the parameter $k$:
\be
\label{k1}
k=-{1\over \gamma^{\prime 2}}\left( {2\alpha^2-1\over 2}\, C'_2 +{6\alpha^2-1\over 12}\, Q_2\right) \, .
\ee
Finally, the quartic terms relate $\xi$ to the integration
constant $C_s$ appearing in $K_s$,
\be
\label{CS}
C_s= {1\over \gamma^{\prime 6}}\left({4\alpha^2-1 \over 2\alpha^2}\, C'_4
+{6\alpha^2-1\over 12\alpha^2}\, Q_4+{2\alpha^2-1\over 4\alpha^2}\,r_4\xi^4- {6\alpha^2-
1\over 4\alpha^2}\, p_4\xi^4\right) \, .
\ee
\\
$~$\\
At this point, our choice of anzats (\ref{anzats}) and constants $\gamma'$, $\xi$ allows to reduce the differential system for
$\Phi_s$, $\Phi$ and the gravity to the last equation. We thus concentrate on the Friedman equation
(\ref{Friedman}) in the background
of the critical  trajectory $\dot\Phi^2=(H^2/\alpha^2)$,
\be
\left({6\alpha^2-1\over 6\alpha^2}\right)~3H^2 =  -{3k\over a^2}+ \rho +{1\over
2}~e^{2\alpha \Phi} \dot \Phi_s^2+V\, .
\ee
The dilatation factor in front of $3H^2$  can be absorbed in
the definition of $\lambda$, $\hat k$ and $C_{R}$, once we take into account eqs. (\ref{C0'}), (\ref{k1}) and (\ref{CS}),
 \be
\label{cosmoeff}
3H^2= 3\lambda-{ 3\hat k\over a^2} +{C_R \over a^4}\, ,
\ee
where
\be
3\lambda=\alpha^2\, Q_0 \, ,
\ee
\be
\hat k={\alpha^2\over \gamma^{\prime 2}}\left( {2\over 6\alpha^2-1}
\, \xi^2 \hat M^2-C'_2 -
{1 \over 2}\, \, Q_2 \right)\, ,
\ee
and
\be
C_R={3\over 2\gamma^{\prime 4}}\left((r_4-p_4)\, \xi^4+2C'_4 +
{1\over 3}\, Q_4\right)\, .
\ee
We note that for $Q_0>0$, $\l$ is positive. In that case, the constraint
(\ref{CS}) allows us to choose a lower bound for the arbitrary constant
$C_s$, so that $\xi^4$ is large enough to have $\hat k> 0$.
This means that the theory is effectively indistinguishable
with that of  a universe with cosmological constant $3\l$,
uniform space curvature $\hat k$, and filled with a thermal
bath of radiation coupled to gravity. This can be easily
seen by considering the Lagrangian
\be
 \label{lag2}
\sqrt{-\det g}
\left[{1\over 2}R -3\l \right]
\ee
and the metric anzats (\ref{metric}), with a 3-space of constant curvature $\hk$. In the action, one can take
into account a uniform space filling bath of massless particles
by adding a Lagrangian density proportional to $1/a^4$
(see \cite{MSSThermal, radiation-density, KPThermal}) in the MSS form. One  finds
\be
\label{EffectiveMSS}
S_{\mbox{\tiny \em{eff}}}=-{|\hk|^{-{3\over 2}}\over 6}\int dt  \, N a^3\left( {3\over N^2}
\left( {{\dot a}\over a}\right)^2+
3\lambda - {3\hat k \over a^2}+{C_R \over a^4}\right) \, ,
\ee
whose variation with respect to $N$ gives (\ref{cosmoeff}).
Actually, the thermal bath interpretation is allowed
as long as $C_R\ge 0$, since the $1/a^4$ term is an
energy density. However, in the case under consideration,
the effective $C_R$ can be negative due to the $m^4$
contribution of the effective potential. The general
solution of the effective MSS action of eq. (\ref{EffectiveMSS})
with $\lambda > 0$, $\hat k  >0$ and $C_R>0$ was recently investigated
in ref.  \cite{KPThermal}. It amounts to a thermally
deformed de Sitter solution, while the pure radiation case
where $\lambda=\hat k=0$, $C_R>0$ was studied in ref. \cite{AKcritical}.
In the latter case,  the time trajectory (\ref{anzats})
 was shown to be an attractor at late times, giving rise to
 a radiation evolving universe with
\be
a= \left( {4C_R\over 3}\right)^{\!\!1/4} \,t^{1/2}\; , ~~~~~~m(\Phi)= {T\over \xi }= {1\over \gamma' a}\, .
\ee
Following ref. \cite{KPThermal}, the general case with $\lambda > 0,
\hat k  >0$ and $C_R>0$ gives rise to cosmological scenarios we summarize here. Depending on the quantity 
\be
\dt={4\over 3} \, {\l\over {\hat k}^2}\, C_R\; , 
\ee
a first or second order phase transition can occur:
\be 
\dt<1\Longleftrightarrow 1^{\mbox{\tiny st}} \mbox{ order transition}\quad , \qquad\dt>1\Longleftrightarrow 2^{\mbox{\tiny d}} \mbox{ order transition}\, .
\ee

\begin{figure}[h!]
\begin{center}
\includegraphics[height=9cm]{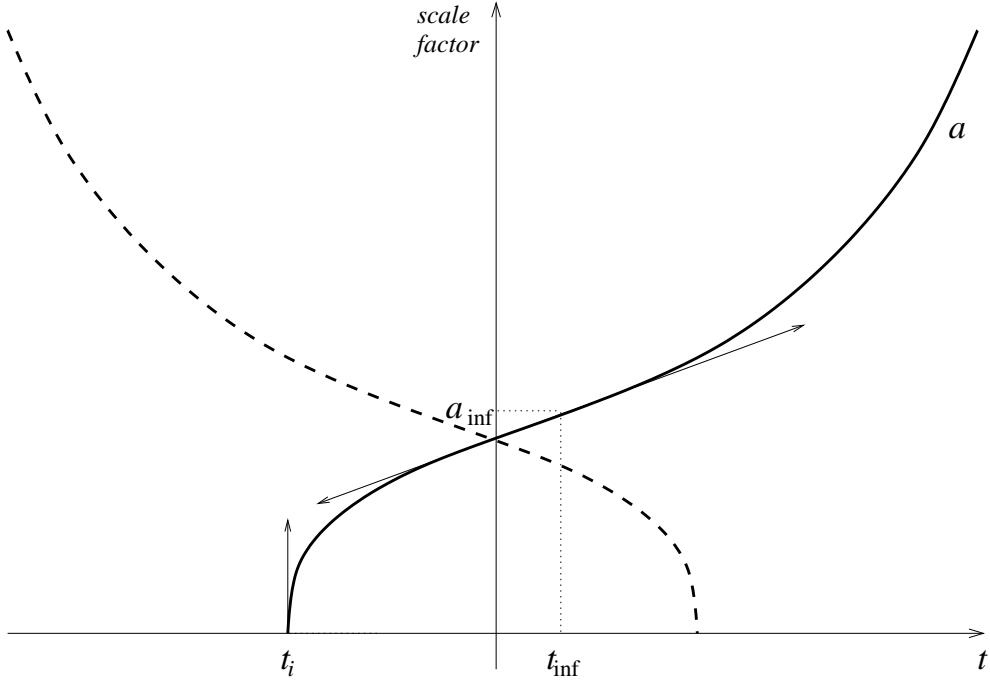}
\caption{\footnotesize \em A second order phase transition occurs. The universe starts with a big bang at $t=t_i$ and evolves smoothly toward the inflationary phase.} 
\label{fig_cosmology_2}
\end{center}
\end{figure}

\begin{figure}[h!]
\begin{center}
\includegraphics[height=9cm]{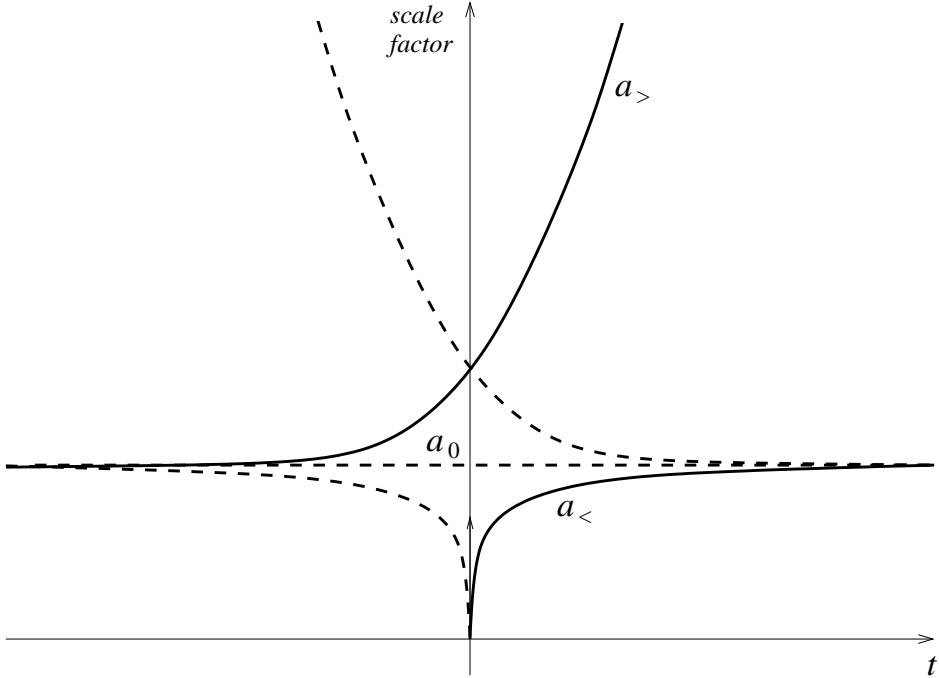}
\caption{\footnotesize \em There are two expanding cosmological evolutions. The first one, $a_{\mbox{\tiny $<$}}$, starts with a big bang and converges quickly to the static solution. The second one, $a_{\mbox{\tiny $>$}}$, is almost static for negative time and inflationary for positive time.} 
\label{fig_cosmology_3}
\end{center}
\end{figure}

\noindent {\large \em i) The case $\dt<1$}\\
There are two cosmological evolutions connected by tunnel effect:
\be
\label{defdeSitt}
a_c(t)=\N  \sqrt{\ve +\cosh^2(\sqrt{\l}\, t)}\; , \;\; t\in \R\, ,
\ee
and
\be
\label{BB}
a_s(t)=\N  \sqrt{\ve -\sinh^2(\sqrt{\l}\, t)}\; , \;\;
t_i\le  t \le -t_i \, ,
\ee
where
\be
\N=\sqrt{{\hat k \over \l}}\, (1-\dt )^{1/4}\; ,
\quad \ve={1\over 2}\left( {1\over \sqrt{1-\dt}}
-1\right)  \; ,
\quad    t_i=-{1\over \sqrt{\l}}\, \mbox{arcsinh}\sqrt \ve\, .
\ee
The ``cosh"-solution  corresponds to a deformation of a standard
de Sitter cosmology, with a contracting phase followed at $t=0$ by
an expanding one. The ``sinh"-solution describes a big bang
with a growing up space till $t=0$, followed  by a contraction
till a big crunch arises. The two evolutions are
connected in Euclidean time by  a {\em $\Phi$-Gravitational Instanton}
\be
\label{inst}
a_E(\tau)=\N \sqrt{\ve + \cos^2 (\sqrt{\l}\, \tau)}\; , \qquad
    \Phi_E (\tau)=-{1\over \alpha}\log\left(\gamma' \,a_E(\tau)\phantom{\dot l}\!\! \right)\, .
\ee
The cosmological scenario is thus starting with  a big bang at  $t_i=-{1\over \sqrt{\l}}\,
\mbox{arcsinh}\sqrt \ve$ and expands up to
 $t=0$, following the ``sinh"-evolution.  At this time, performing 
 the analytic continuation $t=-i(\pi/2\sqrt{\l}+\tau)$
 reaches (\ref{inst}), (where $\tau$ is chosen in the range\footnote{It is also possible to consider the instantons associated to the ranges $\sqrt \l \tau\in [-(2n+1)\pi/2,0]$, $n\in \intN$, see \cite{KPThermal}.}
 $-\pi/2\sqrt{\l}\le \tau\le 0$).  At $\tau=0$,
 a different  analytic continuation to real time exists, $\tau=it$,
 that  gives rise to the inflationary phase of the
 ``cosh"-evolution,  for $t\ge 0$, 
\begin{figure}[h!]
\begin{center}
\includegraphics[height=9cm]{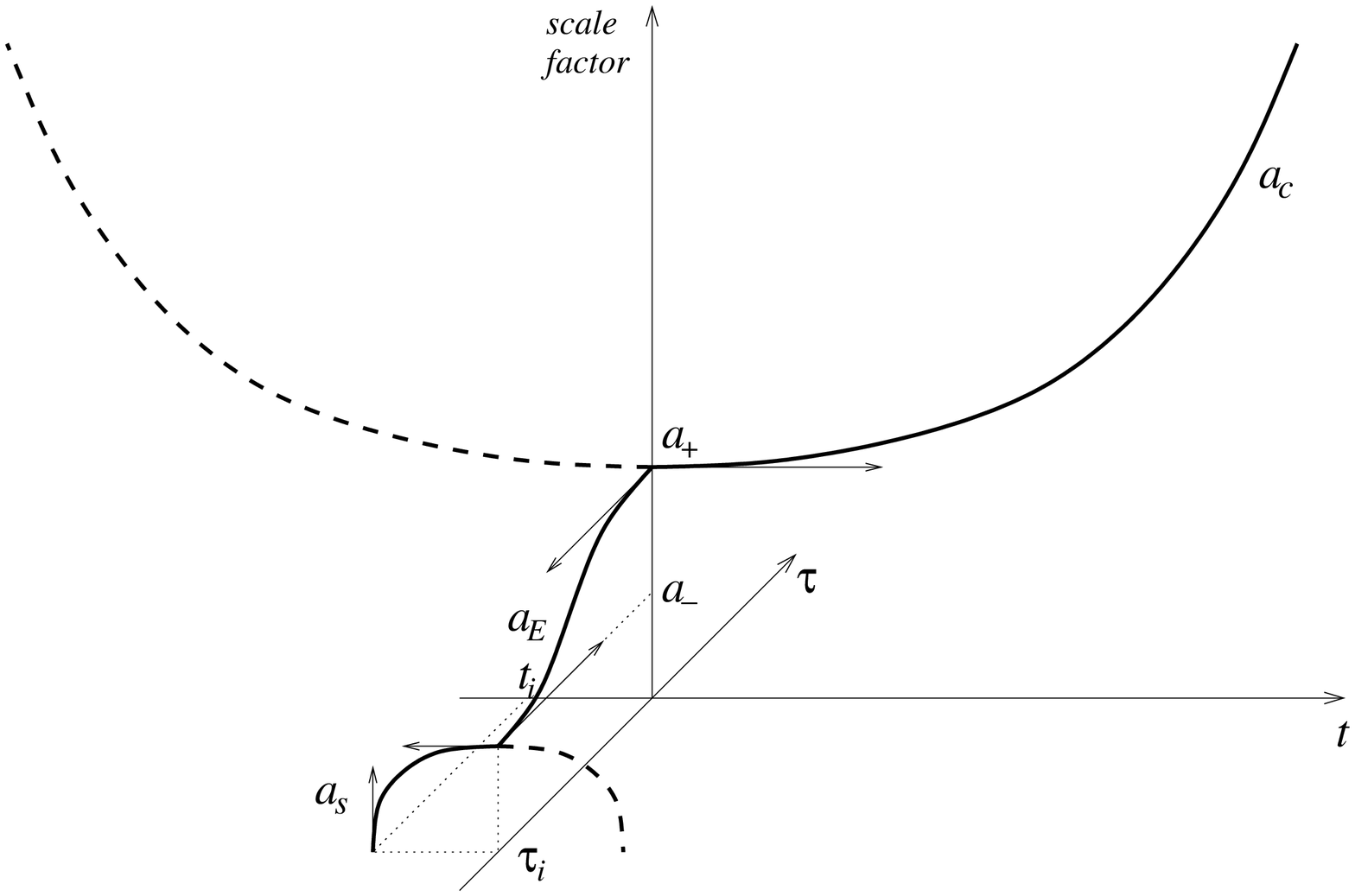}
\caption{\footnotesize \em A first order phase transition can occur. The two cosmological evolutions $a_s$ and
  $a_c$ are connected by an instanton $a_E$. The universe starts with
  a big bang at $t=t_i$ and expands till
  $t=0$ along the solution $a_s$. Then, the scale factor can either contract, or jump
  instantaneously and enter into the inflationary phase
  of $a_c$.} 
\label{fig_cosmology}
\end{center}
\end{figure}
(see fig. \ref{fig_cosmology}). \\
$~$\\
There are thus two possible behaviors when $t=0$ is reached.
Either the universe carries on its ``sinh''-evolution and starts to contract, or a first order transition occurs and the universe enters into the inflationary phase of the 
``cosh''-evolution. In that case, the scale factor  jumps instantaneously   from $a_-$ to $a_+$ at
$t=0$,
\be
a_-=\sqrt{{\hat k\over 2\l}\left(1-\sqrt{1-\dt}\right)} \,\,\,\,
\longrightarrow  \,\,\,\, a_+=\sqrt{{\hat k\over 2\l}\left(1+\sqrt{1-\dt}\right)}\, .
\ee
An estimate of the transition probability is given by
\be
p\propto e^{-2 S_{E\mbox{\tiny \em{eff}}}}\, ,
\ee
where $S_{E\mbox{\tiny \em{eff}}}$ is the Euclidean action
computed with the instanton solution (\ref{inst}), for
$\tau\in[-\pi/2\sqrt{\l}, 0]$. Actually, following
 refs. \cite{radiation-density, KPThermal},
one has:
\be
\label{acT}
S_{E\mbox{\tiny \em{eff}}}=-{1\over 3\l}\sqrt{{1+\sqrt{1-\dt}\over 2}}
\left( E(u)-\left( 1-\sqrt{1-\dt}\right) K(u)\right)\, ,
\ee
where $K$ and $E$ are the complete elliptic integrals of first
and second kind, respectively, and
\be
u= {2(1-\dt)^{1/4}\over \sqrt{1+\sqrt{1-\dt}}}\, .
\ee
$~$\\
{\large \em ii) The case $\dt>1$}\\
There is a cosmological solution,
\be
a(t)=\sqrt{\hk\over 2\l} \, \sqrt{1+\sqrt{\dt-1}\, \sinh(2\sqrt{\l}\, t)}\; , \;\;
t \ge t_i\, ,
\ee
where
\be
t_i=-{1\over 2\sqrt{\l}}\, \mbox{arcsinh}{1\over \sqrt{\dt-1}}\, .
\ee
As in the previous case, it starts with a big bang. However, the behavior evolves toward the inflationary phase in a smooth way, (see fig. \ref{fig_cosmology_2}). 
The transition can be associated to the inflection point arising at $t_{\mbox{\tiny inf}}$, where $a(t_{\mbox{\tiny inf}})=a_{\mbox{\tiny inf}}$,
\be
t_{\mbox{\tiny inf}}={1\over 2\sqrt{\l}}\, \mbox{arcsinh}\sqrt{{\delta_T-1\over \delta_T+1}}\; , \quad a_{\mbox{\tiny inf}}=\sqrt{{\hk\, \delta_T\over 2\l}}\, .
\ee
Another solution, obtained by time reversal $t\to -t$, describes a contracting universe that is ending in a big crunch. \\
$~$\\
{\large \em iii) The case $\dt=1$}\\
There is a static solution, 
\be
a(t)\equiv a_0\quad \mbox{where} \quad a_0=\sqrt{\hk\over 2\l}\, ,
\ee
corresponding to an $S^3$ universe of constant radius. This trivial behavior  can be reached from the cases $i)$ and $ii)$ by taking the limit $\delta_T^2\to 1$.  Beside it, there are two expanding cosmological evolutions,
\be
a_{\mbox{\tiny $<$}}(t)=a_0 \sqrt{1-e^{-2\sqrt{\l}\, t}}\; , \;\; t\ge 0\, ,
\ee
and
\be
a_{\mbox{\tiny $>$}}(t)=a_0 \sqrt{1+e^{2\sqrt{\l}\, t}}\; , \;\; t\in \R\, .
\ee
The first one starts with a big bang at $t=0$, while the second is inflationary. Both are asymtotic to the static one, (see fig. \ref{fig_cosmology_3}). Contracting universes are described by the solutions obtained under the transformation $t\to -t$.
%


\section{Inclusion of Moduli with Other Scaling Properties}
\label{othmod}

\begin{figure}[h!]
\begin{center}
\includegraphics[height=7cm]{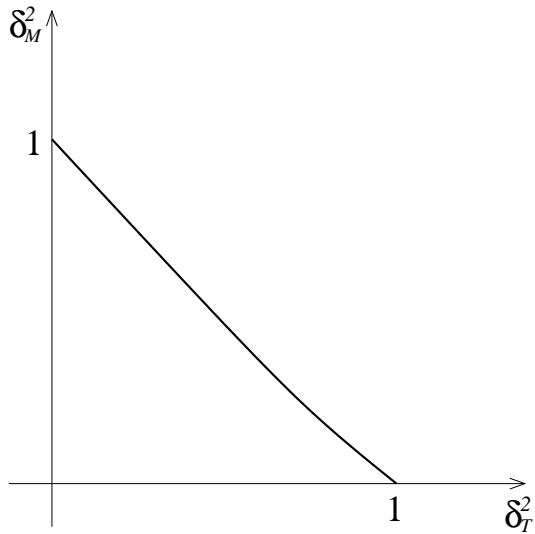}
\caption{\footnotesize \em Phase diagram in the
  $(\dt,\dm)$-plane. Inside the ``almost triangular''
  domain (\ref{domain}), a first order transition can connect two cosmological solutions by tunnel 
  effect. Outside the domain, there is a single expanding solution (and one contracting) that describes a second order transition. On the boundary, there are two expanding (and two contracting) solutions, beside a static one.} 
\label{fig_triangle}
\end{center}
\end{figure}

We would like to consider generalizations of the previous set up.
They are consisting in the inclusion of moduli fields with kinetic
terms obeying  different scaling properties with respect to $\Phi$.
Namely, we take into account the effects of the class of
moduli with Lagrangian density 
\be
\sqrt{- \, \det g}\,  {1\over 2} \, g^{\mu \nu}\,
\left( \partial_\mu \phi_6\partial_\nu \phi_6\,
+\, e^{4\alpha \Phi}\partial_\mu \phi_2\partial_\nu \phi_2\,
+\, e^{6\alpha\Phi}\partial_\mu \phi_0\partial_\nu \phi_0\right)\, ,
\ee
to be added to (\ref{lag}).
With the metric anzats (\ref{metric}), the MSS action
(\ref{act}) is completed by the contributions of the $\phi_w$'s, ($w=0,2,6$),
\be
-{|k|^{-{3\over 2}}\over 6}\int dt \, a^3 \left( -\, {1\over 2N}
\dot \phi_6^2 \, - \, {1\over 2N} e^{4\alpha\Phi}
\dot \phi^2_2\, - \, {1\over 2N} e^{6\alpha\Phi}
\dot \phi^2_0\right)\, .
\ee
The equation of motion for $\Phi$ has now terms arising from $\phi_2$ and $\phi_0$,
\be
\label{phieq}
\ddot\Phi + 3H \dot \Phi +{\partial \over  \partial
\Phi}\left(V-P-{1\over 2}\, e^{2\alpha \Phi} \dot \Phi_s^2 \,
-\, {1\over 2}\,   e^{4\alpha \Phi} \dot \phi^2_2\,-{1\over 2}\,
e^{6\alpha \Phi} \dot \phi^2_0\right)=0\, ,
\ee
while $\Phi_s$ and $\phi_w$ satisfy
\be
\label{moduli}
\ddot \Phi_s + (3H +2 \alpha \dot\Phi )\,\dot\Phi_s =0\; , \; \; \; \; \;\; \; \;
\ddot \phi_w + \left(3H + (6-w) \alpha \dot\Phi \right)\, \dot\phi_w =0 \; .
 \ee
Equations (\ref{moduli}) are trivially solved,
\be
\label{solmod}
K_s\equiv {1\over 2}e^{2\alpha \Phi} \dot\Phi_s^2=
C_{s}{e^{-2\alpha \Phi}\over a^6}\; ,\; \; \; \; \; \;\; \; \;
 K_{w}\equiv {1\over 2}e^{(6-w)\alpha \Phi} \dot \phi^2_w=
 C_{\phi_w}{e^{-(6-w)\alpha \Phi}\over a^6}\; ,
\ee
where $C_s$ and the $C_{\phi_w}$'s are positive constants.
The equivalent equations of motion for $N$ and the scale factor
$a$ have new contributions from the kinetic terms of $\phi_w$, $(w=0,2,6)$.
However, these additional terms cancel out from the linear sum of
the two equations and thus (\ref{gravPotential}) remains invariant. \\
$~$\\
On the critical trajectory (\ref{anzats}), one has
\be
K_s= C_s\, {\gamma^{\prime 2}\over a^4}\; , \; \; \; \; \;\; \; \;  \;
K_w= C_{\phi_w}\, {\gamma^{\prime (6-w)}\over a^w}\;  .
\ee
This implies that the new contributions arising in the $\Phi$
equation (\ref{phieq}) have dimensions two and zero and thus
do not spoil the possible identification between (\ref{phieq})
and the gravity equation (\ref{gravPotential}). In particular,
the $1/a^6$ scaling properties of $\phi_6$ play no role at this stage. The $\Phi$-equation becomes
\be
\begin{array}{lll}
\dot H + 3H^2 =\alpha^2\left(\phantom{\dot\Phi}\!\!\!\!(4C_4'
+Q_4)m^4+ (2C_2'+Q_2) \right. & \!\!\!\!\!\! m^2 & \!\!\!\!\!
+ \, Q_0+(r_4-3p_4)\xi^4m^4  \\ & \!\!\!\!\!\! -& \left.
\!\!\!\!\!\! \!2C_s\gamma^{\prime 6}m^4
-4C_{\phi_2} \gamma^{\prime 6}m^2-6C_{\phi_0} \gamma^{\prime 6}
\phantom{\dot\Phi}\!\!\!\!\right)\, ,
\end{array}
\ee
to be compared with eq. (\ref{gravcri}). The identification of the
constant terms implies
\be
\label{C0bis}
C'_0+6\alpha^2 C_{\phi_0} \gamma^{\prime 6}={6\alpha^2-1\over 6}\, Q_0\, ,
\ee
which is an equation for $\gamma'$. For $Q_0\le 0$, there is always a unique solution for $\gamma'$. However, it is interesting to note that for $Q_0>0$, there is a range for $C_{\phi_0}>0$ where there are always two solutions for $\gamma'$. This case is thus giving rise to two different critical trajectories. The $m^2$ contributions impose
\be
\label{ktot}
k={1\over \gamma^{\prime 2}}\left(2\alpha^2C_{\phi_2} \gamma^{\prime 6}
- {2\alpha^2-1\over 2}\, C'_2 +{6\alpha^2-1\over 12}\,Q_2 \right)\, .
\ee
This fixes the value of $k$ for any arbitrary $C_{\phi_2}$.
Finally, the equation implied by the quartic mass terms is identical
to the one of the previous section, and relates $\xi$ to $C_s$.
We repeat it here for completeness,
\be
\label{CS2}
C_s= {1\over \gamma^{\prime 6}}\left({4\alpha^2-1 \over 2\alpha^2}\,C'_4
+{6\alpha^2-1\over 12\alpha^2}\,Q_4+{2\alpha^2-1\over 4\alpha^2}\,r_4\xi^4-
{6\alpha^2-1\over 4\alpha^2}\,p_4\xi^4\right) \, .
\ee
We would like to stress again that for $C_s$ sufficiently large,
$\xi^4$ can take any value other some bound we may wish.
\\
$~$\\
The Friedman equation in the presence of the extra moduli becomes,
\be
3H^2=  -{3k\over a^2}+ \rho+{1\over2} \dot\Phi^2 +{1\over
2}~ \dot \phi_6^2+{1\over
2}~e^{2\alpha \Phi} \dot \Phi_s^2+{1\over
2}~e^{4\alpha \Phi} \dot \phi^2_2+{1\over
2}~e^{6\alpha \Phi} \dot \phi_0^2 +V\, .
\ee
On the critical trajectory where $\dot\Phi^2=(H^2/\alpha^2)$,
and taking into account eqs. (\ref{C0bis}), (\ref{ktot})
and (\ref{CS2}), one obtains
\be
\label{coseff}
3H^2= 3\lambda-{ 3\hat k\over a^2} +{C_R \over a^4}+{C_M \over a^6}\, ,
\ee
where
\be
3\lambda=\alpha^2\left( Q_0-6\, C_{\phi_0} \gamma^{\prime 6}\right) \, ,
\ee
\be
\label{hk2}
\hat k={\alpha^2\over \gamma^{\prime 2}}\left( 2C_{\phi_2}
\gamma^{\prime 6}+{2\over 6\alpha^2-1}\, \xi^2 \hat M^2-C'_2 -
{1 \over 2}\, Q_2 \right)\, ,
\ee
\be
\label{CR2}
C_R={3\over 2\gamma^{\prime 4}}\left((r_4-p_4)\, \xi^4+2C'_4 
+ {1\over 3}\, Q_4\right)\, ,
\ee
and
\be
C_M={6\alpha^2\over 6\alpha^2-1}\, C_{\phi_6}\, .
\ee
Some observations are in order:\\
{\scriptsize $\bullet$}  $C_{\phi_0}$ gives rise to  a negative contribution
to the cosmological term $3\l$.  \\
{\scriptsize $\bullet$} As previously, it is possible to have $C_R$ positive if one wishes,  by considering large enough values for $\xi^4$. This condition can always be satisfied due to the freedom on $C_s$ in eq. (\ref{CS2}). To reach positive values of  $\hat k$, one can either consider a large enough $\xi^4$ or utilize  $C_{\phi_2}$ as a parameter. \\
{\scriptsize $\bullet$} $C_M$ is always positive and is determined by the modulus  $\phi_6$ only . \\
{\scriptsize $\bullet$}  Here also the system can  be described by
an effective MSS action similar to the one examined
in \cite{KPThermal},
\be
S_{\mbox{\tiny \em{eff}}}=-{|\hk|^{-{3\over 2}}\over 6}\int dt  \, N a^3\left( {3\over N^2}\left( {{\dot a}\over a}\right)^2
+3\lambda - {3\hat k \over a^2}+{C_R \over a^4}+{C_M \over a^6}\right) \, ,
\ee
whose associated Friedman equation is precisely (\ref{coseff}). Thus, once taking into account the thermal and quantum corrections
as well as the effects of the moduli we are considering here,
the system  admits solutions that cannot be distinguished from
the de Sitter cosmology deformed by the presence of thermal
radiation  and time dependent $\phi_6$-moduli fields 
\cite{KPThermal};  this interpretation is  valid only when $C_R$ is positive.\\
$~$\\
Assuming $\l$, $\hat k$, $C_R$ positive  and utilizing the equivalence of the
effective MSS action to the thermally and moduli deformed de
Sitter action studied in \cite{KPThermal}, we can immediately derive
the general solution of the system under investigation.
Our results are summarized as follows (more details can be
found in \cite{KPThermal}).\\
$~$\\
For convenience, we choose rescaled parameters $\dt$, $\dm$,
\be
\dt={4\over 3} \, {\l\over {\hat k}^2}\, C_R\; , \quad \dm={9\over 4} \, {\l^2\over {\hat k}^3}\, C_M\; ,
\ee
and define the domain
\be
\label{domain}
\dt\le1 \quad \mbox{and} \quad \dm\le {1\over 2}\left( \sqrt{1-{3\over 4}\dt}+1\right)^2\left(
  \sqrt{1-{3\over 4}\dt} - {1\over 2} \right)
\ee
in the $(\dt,\dm)$-plane, as shown in fig. \ref{fig_triangle}. 
The Friedman equation (\ref{coseff}) admits solutions that involve a first order phase transition inside this domain and a second order one outside it. It is convenient to express these conditions in terms of $\k$, the only real positive root of the polynomial equation 
\be 
\label{d3}
\k^3+\k^2+{\d^2_T\over 4}\k-{4\over 27}\dm=0 \, ,
\ee
by defining
\be 
\label{Delta}
\D\equiv {4\k^2+4\k+\dt\over (1+\k)^2}={16\over 27}{\dm\over  \k
  (1+\k)^2}\, ,
\ee
where the second equality in eq. (\ref{Delta}) is just a consequence of eq. (\ref{d3}):
\be 
\D<1\Longleftrightarrow 1^{\mbox{\tiny st}} \mbox{ order transition}\quad , \qquad\D>1\Longleftrightarrow 2^{\mbox{\tiny d}} \mbox{ order transition}\, . 
\ee

\noindent{\large \em i) The case $\D<1$}\\
There are two cosmological cosmological solutions connected via tunneling. The first one takes a simple form in terms of a function $\tt(t)$,
\be
\label{defdeSitt2}
a_c(\tt(t))=\N  \sqrt{\ve +\cosh^2(\sqrt{\l}\, \tt(t))}\; ,
\ee
where
\be
\label{epsneg}
 \N = \sqrt{{\hat k (1+\k) \over \l}}\, (1-\D)^{1/4}\; , \quad
 \ve={1\over 2}\left({1\over \sqrt{1-\Delta}}-1\right)\, , 
\ee
and $\tt(t)$ is found by inverting the definition of $t$ as a function of $\tt$,
\be
\label{t1}
t=\int_0^{\tt} dv \sqrt{\cosh^2(\sqrt \l v)+\ve\over \cosh^2(\sqrt \l
  v)+\ve+\tve} \quad \mbox{where}\quad
\tve={\k \over (1+\k)\sqrt{1-\Delta}}\, . 
\ee
In $a_c$, the variables $t$ and $\tt$ are arbitrary in $\R$. The second cosmological evolution is 
\be
\label{BB2}
a_s(\tt(t))=\N  \sqrt{\ve -\sinh^2(\sqrt{\l}\, \tt(t))} \quad \mbox{where} \quad t=-\int^0_{\tt} dv \sqrt{\ve-\sinh^2(\sqrt \l
  v)\over \ve+\tve-\sinh^2(\sqrt \l v)}\, ,
\end{equation}
with the range of time 
\be 
\label{ti}
\tt_i\equiv-{1\over \sqrt{\lambda}}\mbox{arcsinh}\sqrt{\varepsilon} \le\tt \le -\tt_i\quad
\mbox{\ie}\quad t_i=-\int^0_{\tt_i} dv
\sqrt{\ve-\sinh^2(\sqrt \l v)\over
  \ve+\tve-\sinh^2(\sqrt \l v)}\le t\le -t_i\, . 
\ee
As in the previous section, the ``cosh"-solution describes a contracting phase followed by
an expanding one and approaches a standard de Sitter cosmology for positive or negative large times. The ``sinh"-solution starts with a big bang, ends with a big crunch, while the scale factor reaches its maximum at $t=0$. The two cosmological solutions are
connected by  a {\em $\Phi$-Gravitational Instanton}
\be
\label{inst2}
a_E(\ttau(\tau))=\N \sqrt{\ve + \cos^2 (\sqrt{\l}\, \ttau(\tau))}\; , \qquad
    \Phi_E (\ttau(\tau))=-{1\over \alpha}\log\left(\gamma' \,a_E(\ttau(\tau))\phantom{\dot l}\!\! \right)\, ,
\ee
where
$\ttau(\tau)$ is the inverse function of
\be
 \t=-\int^0_{\taut} dv \sqrt{\cos^2(\sqrt \l
  v)+\ve\over \cos^2(\sqrt \l v)+\ve+\tve}\, ,
\ee
and the range of Euclidean time is\footnote{Actually, one can also consider the instantons associated to the ranges $\sqrt \l \ttau\in [-(2n+1)\pi/2,0]$, $n\in \intN$, see \cite{KPThermal}.} 
\be
-{\pi\over 2\sqrt \l}\le   \taut \le 0 \quad \mbox{\ie} \quad \t_i \le  \t\le 0 \quad  \mbox{where} \quad \t_i=
-\int^0_{-{\pi\over 2\sqrt\l}} dv \sqrt{\cos^2(\sqrt \l v)+\ve\over
  \cos^2(\sqrt \l v)+\ve+\tve}.
\ee
The cosmological scenario starts with an initial singularity at $t_i$ and follows the ``sinh''-expansion till $t=0$. At this time, the solution can be analytically continued to the instantonic one by choosing $\tt=-i(\pi/2\sqrt{\l}+\ttau)$ \ie $t=-i(-\tau_i+\tau)$. When the Euclidean time $\tau=0$ is reached, a second analytic continuation to real time, $\ttau=i\tt$ \ie $\tau=it$, gives rise to the inflationary phase of the ``cosh''-solution, for later times $t\ge 0$, (see fig. \ref{fig_cosmology}). 
At $t=0$, the universe has thus two different possible behaviors. It can carry on its evolution along  the ``sinh''-solution \ie enter in a phase of contraction. Or, a first order phase transition occurs and the trajectory switches to the ``cosh''-evolution. In that case, the scale factor jumps instantaneously from $a_-$ to $a_+$,
 \be
a_-=\sqrt{{\hat k(1+\k) \over 2\l}\left(1-\sqrt{1-\D}\right)} \,\,\,\,
\longrightarrow  \,\,\,\, a_+=\sqrt{{\hat k (1+\k) \over 2\l}\left(1+\sqrt{1-\D}\right)}\, .
\ee
The transition probability is controlled by the Euclidean action, $p\propto e^{-2 S_{E\mbox{\tiny \em{eff}}}}$, where $S_{E\mbox{\tiny \em{eff}}}$ has been computed in \cite{KPThermal}. For
$\sqrt \l \ttau\in[-\pi/2, 0]$, one has
\be
S_{E\mbox{\tiny \em{eff}}}=-{1\over 3^{5/4}\l}\left(4-3\dt\right)^{1/4}\sqrt{\sin\left({\theta+\pi\over 3}\right)}\left( E(u)- {\sqrt{4-3\dt}\cos\left({\theta\over 3}\right)-1+{3\over 2}\dt \over \sqrt{3}\sqrt{4-3\dt}\sin\left({\theta +\pi\over 3}\right)} K(u)\right)\, ,
\ee
where 
\be
u=\sqrt{{\sin\left({\theta \over 3}\right)\over \sin\left({\theta +\pi\over 3}\right)}}\; ,\qquad \theta= \arccos\left({16\dm+9\dt-8\over \left(4-3\dt\right)^{3/2}}\right)\, .
\ee\\
$~$\\
We have displayed the scale factors $a_c$ and $a_s$ in terms of $\dt$ and $\dm$. However, it is interesting to express the solutions with more intuitive quantities, namely the temperatures $T_+$ at $t=0_+$ (along the ``cosh''-evolution) and $T_-$ at $t=0_-$ (along the ``sinh''-evolution). Using the fact that $a_\pm T_\pm = \xi/\gamma'$, one finds
\be
\label{T+-}
T_\pm (\dt,\dm)=T_m {\sqrt{\delta_T}\over
  \sqrt{(1+\k)\left(1\pm\sqrt{1-\D}\right)}}\quad \mbox{where}\quad 
T_m={\xi\over \gamma'}\sqrt{{2\l\over \hk \delta_T}}\, .
\ee
It is shown in \cite{KPThermal} that the cosmological solutions can be written as
\be
\label{AC}
a_c(\tt(t))={\sqrt{\hk} \over\sqrt{2\l(1-\A)}}\sqrt{1+{T_-^2-T_+^2\over T_-^2+T_+^2}\cosh 
  (2\sqrt\l \,\tt(t))}\; , \quad t > 0 
\ee
and
\be
\label{AS}
a_s(\tt(t))={\sqrt{\hk} \over\sqrt{2\l(1-\A)}}\sqrt{1+{T_+^2-T_-^2\over T_+^4+T_-^2}\cosh 
  (2\sqrt\l\, \tt(t))}\; , \quad t < 0\, , 
\ee
where we have defined
\be
\A={1-{T^2_+T^2_-/T^4_m}\over \left(T_+/T_-+T_-/T_+\right)^2}\, .
\ee 
We note that there is  a temperature duality $T_+ \leftrightarrow T_-$ that switches the two cosmological solutions $a_c(\tt(t))$ for  $t>0$ and $a_s(\tt(t))$ for $t<0$ into each other:
\be
\left(\,T_+ \leftrightarrow T_-\right) \quad  \Longleftrightarrow \quad \left(a_c(\tt(t))\; \mbox{for} \;  t >0   \leftrightarrow a_s(\tt(t))\; \mbox{for}\; t<0 \phantom{\dot \Phi}\!\!\!\right)\, . 
\ee 
$~$\\
Once quantum and thermal corrections are taken into account, the no-scale supergravities we are considering share a common effective behavior with the thermally and moduli deformed de Sitter evolution (see eq. (\ref{coseff})). This means that the temperature $T_m$ can be defined in both contexts. In eq. (\ref{T+-}), $T_m$ is expressed in terms of critical trajectory quantities. However, one can consider the effective $1/a^4$-radiation energy density in eq. (\ref{coseff}) to define the ``would-be number" of massless bosonic (fermionic) degrees of freedom, $n_{\mbox{\tiny \em{eff}}}^{B(F)}$, of the equivalent deformed de Sitter point of view:
\be
\label{u}
\rho_R \equiv {C_R\over a^4} \equiv  {\pi^4\over 15}\left(n_{\mbox{\tiny \em{eff}}}^B + {7\over 8} n_{\mbox{\tiny \em{eff}}}^F\right) T^4\, ,
\ee
where we have applied the relations (\ref{rP}) and (\ref{i}) for massless states. Using eq. (\ref{u}) and the fact that $aT\equiv a_\pm T_\pm$, one can rewrite $T_m$ as 
\be
T_m= \left({45\over \pi^4}\; {\l \over \left.n_{\mbox{\tiny \em{eff}}}^B + {7\over 8} n_{\mbox{\tiny \em{eff}}}^F\right.}\right)^{1/4}\, .
\ee
$~$\\
{\large \em ii) The case $\D>1$}\\
There is an expanding solution we briefly describe (a contracting one is obtained by time reversal $t\to -t$),
\be
a(\tt(t))=\sqrt{\hk(1+\k) \over 2\l} \, \sqrt{1+\sqrt{\D-1}\, \sinh(2\sqrt{\l}\, \tt(t))}\, ,
\ee
where $t$ as a function of $\tt$ is given by
\be
t=\int_0^{\tt} dv \sqrt{\sqrt{\D-1}\, \sinh(2\sqrt \l
  v)+1\over \sqrt{\D-1}\, \sinh(\sqrt 2\l v)+1+{2\k\over 1+\k}}
 \ee
and we consider
\be
\tt\ge \tt_i\equiv -{1\over 2\sqrt{\l}}\, \mbox{arcsinh}{1\over \sqrt{\D-1}}\quad \mbox{\ie} \quad t\ge t_i\equiv -\int^0_{\tt_i} dv \sqrt{\sqrt{\D-1}\, \sinh(2\sqrt \l
  v)+1\over \sqrt{\D-1}\, \sinh(\sqrt 2\l v)+1+{2\k\over 1+\k}}\, .
\ee
This cosmological solution describes a smooth evolution from a big bang to an inflationary era, (see fig. \ref{fig_cosmology_2}). It has a single inflection point  arising when $a=a_{\mbox{\tiny inf}}$, which is  defined by the following condition,
\be
{\l a_{\mbox{\tiny inf}}^2 \over \hk^2}>0 \quad \mbox{satisfies} \quad x^3-{\dt\over 4}x-{8\over 27}\dm=0\, .
\ee 
$~$\\
{\large \em iii) The case $\D=1$}\\
Beside the following static solution, 
\be
\label{Static}
a(t)\equiv a_0 \quad \mbox{where}\quad a_0=\sqrt{{\hk(1+\k) \over 2\l}}\equiv \sqrt{{\hk\over 3\l}\left({1+\sqrt{1-{3\over 4}\dt}}\right)}\, ,
\ee
the two expanding cosmological evolutions we found for vanishing $\dm$ are generalized by,
\be
a_{\mbox{\tiny $<$}}(\tt(t))=a_0 \sqrt{1- e^{-2\sqrt{\l}\, \tt(t)}}\quad \mbox{where}  \quad t=\int_0^{\tt}dv \sqrt{{1-e^{-2\sqrt{\l}v}\over 1+{2\k'\over 1+\k'}-e^{-2\sqrt{\l}v}}}\, ,
\ee
for $\tt\ge0$ \ie $t\ge0$, and
\be
a_{\mbox{\tiny $>$}}(\tt(t))=a_0 \sqrt{1+ e^{2\sqrt{\l}\, \tt(t)}}\quad \mbox{where} \quad t=\int_0^{\tt}dv \sqrt{{1+e^{2\sqrt{\l}v}\over 1+{2\k'\over 1+\k'}+e^{2\sqrt{\l}v}}}\, ,
\ee
for arbitrary $\tt$ and $t$. They are monotonically increasing as in the pure thermal case: the first one starts with a big bang and the second one is inflationary, (see fig. \ref{fig_cosmology_3}). Two other solutions obtained under $t\to-t$ are decreasing. These time-dependent solutions are asymtptic to the static one.\\ 
$~$\\
Before concluding
let us signal that the number $\hn_0^{B(F)}$ of states  with $\Phi$-independent
masses below $T(t)$ is not strictly speaking a constant. It is
actually lowered by one unit each time the temperature $T(t)$ passes
below the mass threshold $m_{\hi}$ of a boson (fermion). Our critical solutions
for the scale factor are thus well defined in any range of time where $\hn_0^{B}$ and $\hn_0^{F}$
are constant. The full cosmological scenario is then
obtained by gluing one after another these ranges. Each time a mass 
threshold is passed, the values of $\hn_0^{B(F)}$, $\hat M^2$, $r_4$ and $p_4$
decrease (see eqs. (\ref{T2}), (\ref{r4c}) and (\ref{P4c})), and the
parameters of the critical trajectory have to be evaluated
again. However, the constraint (\ref{CS2}) implies that
$\left(\hn_0^{B}+{7\over 8}\hn_0^{F}\right) \xi^4$
{\em remains constant}, due to the fact that $C_s$ (and any other
$C_{\phi_w}$) is constant along the full cosmological
evolution. This implies that $C_R$ defined in eq. (\ref{CR2}) is also
invariant.   
However, the term $\xi^2 \hat M^2$ and thus  $\hk$
can decrease (see eq. (\ref{hk2})). The
positivity of $\hk$ can nevertheless be guaranteed by the modulus term
$C_{\phi_2}\gamma^6$. Actually, this procedure that is consisting in
gluing time intervals as soon as an energy threshold is reached is
identical to what is assumed in Standard Cosmology. In the latter
case, the full
time evolution is divided in different phases (e.g. radiation
dominated, matter dominated, and so on).


\section{String Perspectives and Conclusions}
\label{perscl}

At the classical string level, it is well known that it is difficult
to construct exact cosmological string solutions. It is even more
difficult to obtain de Sitter like inflationary evolutions, even in
less than four dimensions.\\
$~$\\
The main difficulty comes from the fluxes and torsion terms which are 
created via non-trivial field strength (kinetic terms) and have
the tendency to provide negative contributions to the cosmological
term, thus anti de Sitter like vacua. To illustrate a relative issue, consider for instance the Euclidean version of the de Sitter space
in three dimensions, $dS^3$, which is nothing but the 3-sphere $S^3$.
Although $S^3$ can be represented by  an exact
conformal field theory based on an $SU(2)_k$ WZW model, the latter does not admit
any analytic continuation to real time. This is due to the existence of a
non-trivial torsion $H_{\mu\nu\rho}$ that becomes imaginary under an
analytic continuation \cite{ABS, Townsend2Sonner, BKOP}.
\\
$~$\\
This obstruction in string and M-theory is generic and follows from
the kinetic origin of the flux terms. A way to bypass this fact is to take into account higher derivative terms \cite{deSitterR4}. Another strategy is to  assume non-trivial effective fluxes coming from
negative tension objects such as orientifolds. This idea was
explored in the field theory approximation in ref. \cite{CCK}. To go further, it is necessary to work with string cosmological backgrounds based on exact conformal field
theories. However, the only known exact cosmological solution without
the torsion problem described above is that of
$SL(2,\R)/U(1)_{-|k|}\times K$, \cite{KL}. Its Euclidean version is
also well defined by the parafermionic T-fold \cite{Parafermions, KTT}.\\
$~$\\
In this work we have implemented a more revolutionary approach. We start
with a classical superstring background with spontaneously broken
$N=1$ supersymmetry defined on a flat space-time. The
effective field theories associated to these cases are  nothing but the
$N=1$ string induced ``no-scale supergravity models". Working at the
field theory level, we have shown that the quantum and thermal
corrections create dynamically universal effective potential terms that
give rise to non-trivial cosmological accelerating solutions.\\
$~$\\
The main ingredients we have used are the scaling
properties of the effective potential at finite temperature in the 
``no-scale  $N=1$ spontaneously broken supergravities", once the backgrounds follow critical trajectories where all fundamental scales have a similar evolution in time. Namely, the
supersymmetry breaking scale  $m(\Phi)$, the inverse of the scale factor $a$, the temperature $T$ an the infrared scale $\mu$ remain proportional to each other:
$$
e^{\alpha \Phi}\equiv m(\Phi)\, = \, {1\over\gamma' a}\, = \, {T\over \xi}\, =\, {\gamma\over \gamma'} \, \mu\, .
$$
The ``no-scale modulus $\Phi$" is very special in the sense that it is
the superpartner of the goldstino and couples to the  trace
of the energy momentum of a sub-sector of the theory. It also provides
non-trivial dependences in  the  kinetic terms of other special
moduli of the type:
$$
K_w\equiv -{1\over 2}~e^{(6-w)\alpha\Phi}\, \left(\partial\phi_w\right)^2\, ,\quad (w=0,2,4,6)\, ,
$$
where $\phi_4\equiv \Phi_s$ in the text. 
The quantum and thermal corrections, together with the non-trivial motion of the special moduli, allow to find thermally and moduli deformed de Sitter evolutions. The cosmological
term $3\l(am)$, the curvature term $\hat k(am, a T)$ and the
radiation term $C_R(am, a T)$ (see eqs. (\ref{cosmoeff}) or (\ref{coseff})), are dynamically generated in a
\emph{controllable way} and are effectively constant. 
Obviously, as stated in the introduction, these solutions are valid below Hagedorn-like scales associated to the temperature as well as the supersymmetry breaking scale $m$,  where instabilities would occur in the extension of our work in a stringy framework. These restrictions on $m$ are supported by the analysis of the string theory examples considered in ref. \cite{CKPT}.    
\\
$~$\\
When the deformation of the de Sitter evolution is below some critical value, there are two cosmological solutions which are
connected by tunnel effect and interchanged under a temperature duality.  The first one describes a big bang with a
growing up space till  $t=0$, followed  by a contraction that ends with  a big
crunch. The second corresponds to a deformation of a standard
de Sitter evolution, with a contracting phase followed at  $t=0$ by
an expanding one. The universe starts on the big bang cosmological
branch and expands up to
 $t=0$ along the first solution. At this time, two distinct behaviors can occur. Either the universe starts to contract, or a
first order phase transition arises via a $\Phi$-gravitational instanton, toward the
inflationary phase of the deformed de Sitter evolution. The transition probability $p$ can be estimated. \\
$~$\\
If on the contrary the induced cosmology corresponds to a de Sitter-like  universe with deformation above the critical value, the previous big bang and inflationary behaviors are smoothly connected via a second order phase transition. \\
$~$\\
It is of main importance that the field theory approach we developed
here can easily be adapted at the string level \cite{CKPT}, following the recent
progress in understanding the stringy wave-function of the universe 
\cite{KTT, StringWF}. This will
permit us to go beyond the
Hagedorn temperature and understand better the very early ``stringy"
phase  of our universe \cite{HagedornPhase}.
At this point, we insist again on a highly interesting question that can be raised concerning the common wisdom which states that all radii-like moduli should be large to avoid Hagedorn-like instabilities. If this statement was true, then the quantum and thermal corrections should be considered in a 10 rather than in a 4 dimensional picture. However, the recent results of ref. \cite{CKPT}  where explicit  string models are considered show that this is only valid for the radii-moduli which are participating to the supersymmetry breaking mechanism. The remaining ones are free from any Hagedorn-like instabilities and can take very small values, even of the order of the string scale. 


\section*{Acknowledgements}

We are grateful to Constantin Bachas, Ioannis Bakas, Adel Bilal, Gary Gibbons, John Iliopoulos, Jan Troost and Nicolas Toumbas for  discussions. \\
\noindent
The work of C.K. and H.P. is partially supported by the EU contract MRTN-CT-2004-005104 and the ANR (CNRS-USAR) contract  05-BLAN-0079-01. C.K. is also supported by the UE contract MRTN-CT-2004-512194, while H.P. is supported by the UE contracts MRTN-CT-2004-503369 and MEXT-CT-2003-509661, INTAS grant 03-51-6346, and CNRS PICS 2530, 3059 and 3747.


\end{document}